\RequirePackage{fix-cm}
\documentclass[smallextended]{svjour3}       
\pdfoutput=1
\smartqed  
\usepackage{url}
\usepackage{amsmath}
\usepackage{amssymb}
\usepackage[T1]{fontenc}
\usepackage[utf8]{inputenc}
\RequirePackage{mathptmx}
\usepackage{graphicx}

\usepackage{latexsym}

\begin{document}

\title{Materials challenges for quantum technologies based on color centers in diamond
}


\author{Lila V. H. Rodgers \and Lillian B. Hughes \and Mouzhe Xie \and Peter C. Maurer \and Shimon Kolkowitz \and Ania C. Bleszynski Jayich \and Nathalie P. de Leon
}


\institute{ L. V. H. Rodgers \at
            Department of Electrical Engineering, Princeton University, Princeton, New Jersey 08544, USA \\
            \email{lvhr@princeton.edu}
            \and
            L. B. Hughes \at
            Materials Department, University of California Santa Barbara, Santa Barbara, California 93106, USA \\
            \email{lbhughes@ucsb.edu}
            \and
            M. Xie \at
            Pritzker School of Molecular Engineering, University of Chicago, Chicago, Illinois, USA \\
            \email{xiemouzhe@uchicago.edu}
            \and
            P. Maurer \at
            Pritzker School of Molecular Engineering, University of Chicago, Chicago, Illinois, USA \\
            \email{pmaurer@uchicago.edu}
            \and
            S. Kolkowitz \at
            Department of Physics, University of Wisconsin-Madison, Madison, Wisconsin 53706,USA \\
            \email{kolkowitz@wisc.edu}
            \and
            A. C. Bleszynski Jayich \at
            Department of Physics, University of California Santa Barbara, Santa Barbara, California 93106, USA \\
            \email{ania@physics.ucsb.edu}
            \and
            N. P. de Leon \at
            Department of Electrical Engineering, Princeton University, Princeton, New Jersey 08544, USA \\
            \email{npdeleon@princeton.edu}           
}


\maketitle

\begin{abstract}
Emerging quantum technologies require precise control over quantum systems of increasing complexity. Defects in diamond, particularly the negatively charged nitrogen-vacancy (NV) center, are a promising platform with the potential to enable technologies ranging from ultra-sensitive nanoscale quantum sensors, to quantum repeaters for long distance quantum networks, to simulators of complex dynamical processes in many-body quantum systems, to scalable quantum computers. While these advances are due in large part to the distinct material properties of diamond, the uniqueness of this material also presents difficulties, and there is a growing need for novel materials science techniques for characterization, growth, defect control, and fabrication dedicated to realizing quantum applications with diamond. In this review we identify and discuss the major materials science challenges and opportunities associated with diamond quantum technologies.

\end{abstract}

\maketitle

\section{Introduction} 
Devices that fully harness the laws of quantum mechanics offer the potential to fundamentally revolutionize technologies in diverse areas including sensing, communication, and computation. A key challenge in realizing robust, reliable, and scalable quantum technologies is engineering controlled interactions amongst qubits, or between qubits and classical control hardware, while mitigating uncontrolled interactions with the environment. For solid-state qubits, materials science and engineering will play an important role in addressing this and other challenges related to scaling. 

Color centers in the solid state feature desirable atom-like properties such as exceptional spin coherence and optical addressability, while offering the potential to be integrated into compact devices and scalable technologies. Diamond is a particularly attractive material for quantum technologies: it is a wide band gap semiconductor that is optically transparent, can be composed of stable nuclear-spin-free isotopes, has an exceptionally high (2250 K) Debye temperature, and plays host to many energetically deep defects that can form qubits. In a sense, these properties make diamond a magnetically, electrically, and mechanically `quiet' environment, closely emulating a vacuum. 

Several of diamond's defects have been investigated for quantum applications (Fig.~\ref{fig:Fig1}a,c), most notably the negatively charged NV center \cite{Gruber1997,DOHERTY20131,childress2013}, as well as other defects including SiV$^-$, SiV$^0$, NV$^0$, and N$_S^0$ (substitutional nitrogen) \cite{Bhaskar2020,Rose2018a,Baier2020,Epstein:2005ua,degen2020entanglement}. The NV center is an optically-addressable electronic spin (Fig.~\ref{fig:Fig1}d-g) exhibiting long (millisecond-scale) coherence times at room temperature. Notable achievements with NV quantum sensors include the measurement of transport and magnetic properties in condensed matter systems \cite{Casola2018}, the composition of geological samples \cite{fu2014}, and the imaging and control of complex biological systems \cite{schirhagl2014}. In quantum communication, NV centers have enabled fundamental tests of quantum mechanics \cite{Hensen2015}, and key building blocks of quantum repeaters based on NV and SiV$^-$ centers \cite{Bhaskar2020,Pfaff2014} have been demonstrated. In quantum computation and simulation  NV centers have been used to study many-body dynamics \cite{Choi2017} and to demonstrate various algorithms \cite{van2012decoherence,waldherr2014quantum}. All of these applications are interrelated, and developments in one area will often lead to improvements in another. For example, NV centers can be used as sensors to probe sources of materials-induced decoherence in solid-state quantum computing platforms. In this review we take a holistic view of quantum applications and discuss the key figures of merit (depicted in Fig.~\ref{fig:Fig2}), summarize the current experimental materials challenges in diamond, and envision new opportunities for materials science to advance quantum technologies.

\section{Applications and metrics}
\label{sec:Applications}
\subsection{Quantum sensing} 
\label{sec:Sensing}

\begin{figure*}

\includegraphics[width=174mm]{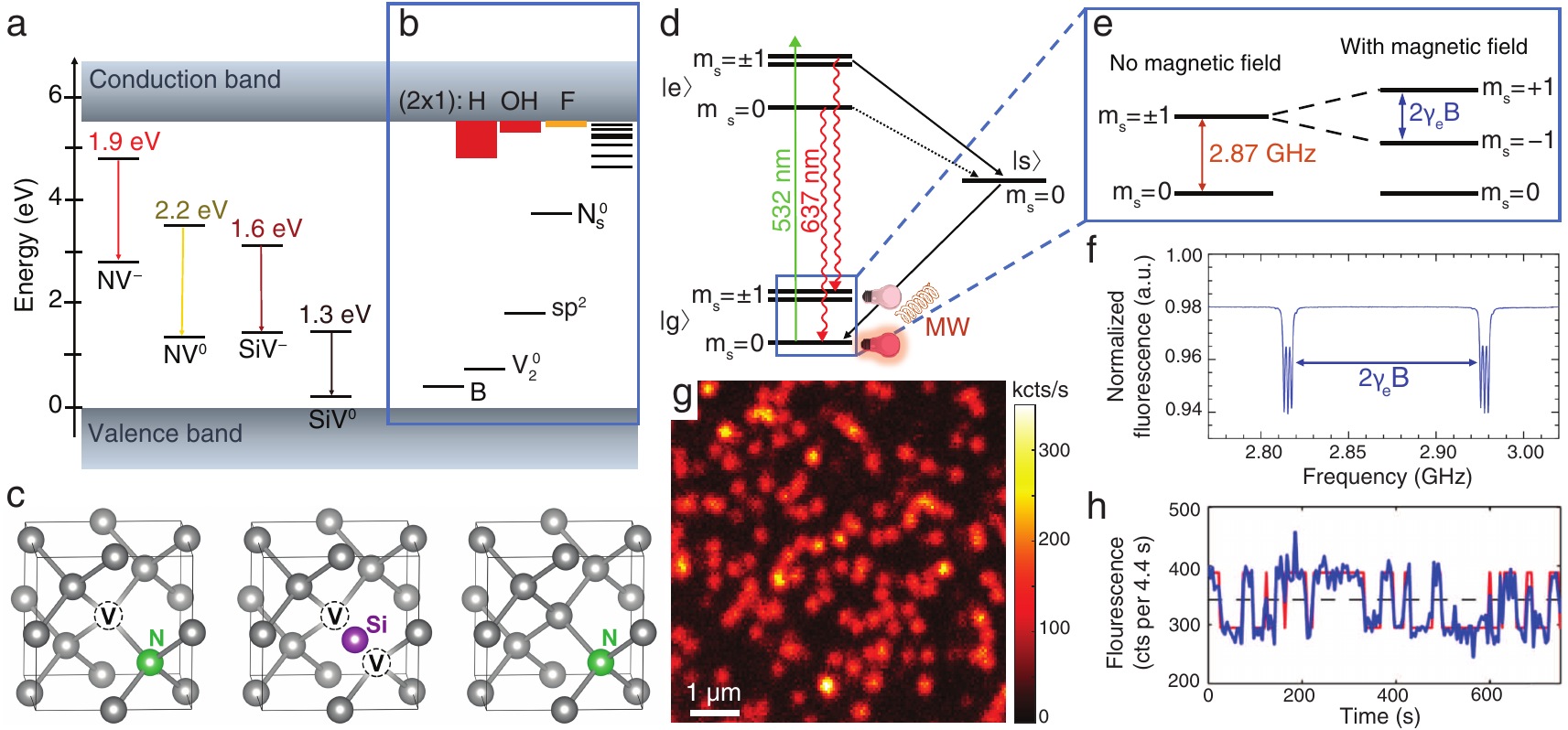}
\caption{\textbf{(a)} Band diagram showing the energies of NV and SiV centers relative to the valence and conduction band edges of diamond. Ground and excited states associated with the most prominent defect zero phonon lines are shown. Ground state positions are estimated from adiabatic charge transition levels as calculated in Ref.~\cite{Deak2014,Thiering2018} and align with experimental photoconductivity measurements to within ± 0.1 eV \cite{Aslam2013,Allers1995}. \textbf{(b)} Energetic positions of common defects in diamond, including boron (B), divacancy (V$_2^0$), sp$^2$, and substitutional nitrogen (N$_s^0$). Levels for deep defects shown are predicted or measured \cite{stacey2019,Deak2014,gheeraert1999}. Surface image states (red) and acceptor states (orange) for H, OH, and F terminations on (100) surfaces with (2x1) reconstruction predicted by density functional theory \cite{kaviani2014}. Black lines on the right indicate continuum of unoccupied states associated with surface defects, as has been observed by near-edge X-ray absorption fine structure (NEXAFS) \cite{Sangtawesin2018,stacey2019}. \textbf{(c)} Representative atomistic diagrams of the defects with carbon (gray), substitutional nitrogen (green), vacancies (white), and interstitial silicon (purple) atoms shown within a diamond unit cell. 
\textbf{(d)} Simplified energy level structure of the NV center showing the spin triplet orbital ground ($|g\rangle$) and excited ($|e\rangle$) states (left) and the metastable singlet state ($|s\rangle$) (right). Red arrows indicate radiative transitions and black arrows illustrate the strong (solid) and weak (dashed) nonradiative intersystem crossings which enable optical polarization and readout of the $m_s$ sublevels in $|g\rangle$. \textbf{(e)} The ground state exhibits a zero-field splitting of 2.87 GHz between $m_s=0$ and $m_s=\pm1$. Application of an external magnetic field $B$ gives rise to Zeeman splitting between the $m_s=\pm1$ states, forming the basis for qubit operation and magnetic field sensing. \textbf{(f)} Continuous wave optically detected ESR spectra show changes in fluorescence when driving the $m_s=0\rightarrow\pm1$ transitions, each of which exhibit additional splitting arising from the hyperfine interaction with the nitrogen nuclear spin (image adapted from Ref. \cite{Barry2020}). \textbf{(g)} Scanning confocal microscope image of a bulk diamond showing fluorescence from optically resolved single NV centers. \textbf{(h)} Fluorescence time trace from Ramsey measurements performed on a single NV center in Ref.~\cite{Maurer2012} showing quantum jumps of a nearby $^{13}$C nuclear spin.}
\label{fig:Fig1}
\end{figure*}

\begin{figure*}

\includegraphics[width=84mm]{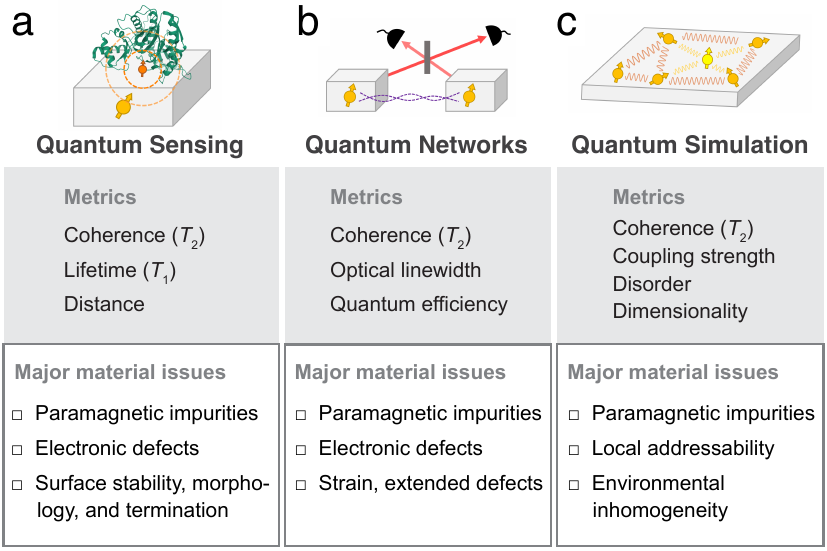}
\caption{Cartoon illustrations and summaries of key metrics and corresponding material issues for quantum applications of diamond color centers. 
}
\label{fig:Fig2}
\end{figure*}

Through their coupling to the surrounding environment, color centers in diamond can be used as local sensors \cite{Barry2020,schirhagl2014}. For example, the electronic spin states of the NV center couple to the magnetic field \cite{maze2008,balasubramanian2008}, electric field \cite{dolde2011}, strain \cite{ovartchaiyapong2014}, and temperature \cite{Kucsko2013} at the location of the defect. This manifests as relative shifts in the energies of the spin sublevels and changes in the transition rates between them, which can be measured through a combination of optical spin polarization, coherent microwave control, and optical readout (Fig.~\ref{fig:Fig1}d-h). 

Quantum sensing with NV centers applies the same tools and techniques originally developed for nuclear magnetic resonance (NMR) and magnetic resonance imaging (MRI). Many of the figures of merit and terminology are therefore the same, such as the spin state lifetime ($T_1$), the spin dephasing time determined by inhomogeneous broadening ($T_2^*$), and the spin coherence time for a given coherent control sequence ($T_2$). These time scales effectively parameterize the noise floor for a given signal of interest due to unwanted noise from the surrounding environment, and therefore play an important role in determining sensitivity. The $T_1$ time is limited by coupling between the spin and phonons in the lattice. Diamond is the stiffest material and has low spin-orbit coupling, allowing NV centers to have exceptionally long $T_1$ at room temperature in the absence of other sources of magnetic or electric field noise. $T_2$ is ultimately limited by $T_1$, and can be shorter because of magnetic noise arising from nuclear spins, paramagnetic atoms, and electronic defect states (Fig.~\ref{fig:Fig1}b, Fig.~\ref{fig:growthfabfig}b). 

Two advantages of NV center-based sensing over conventional NMR/MRI are the higher spatial resolution enabled by individual solid-state defects and the ability to work at weak magnetic fields due to the high degree of spin-polarization afforded by optical spin-pumping. This results in new figures of merit such as the optical spin contrast parameter $C$, a unit-less number from $0-1$ that quantifies the spin readout fidelity of a single measurement. The minimum detectable signal $\delta S$ for a total averaging time $\tau$, coherent duration of a single measurement $T_x$, linear rate of spin state population change per unit target signal $\gamma$ performed using $N$ NVs can be written as

\begin{equation}
    \delta S(\tau) \approx \frac{1}{\gamma C \sqrt{N T_x \tau}}.
    \label{Eq:Sensitivity}
\end{equation}

\noindent As a concrete example, consider the optically detected magnetic resonance (ODMR) spectrum from a single NV center shown in Fig.~\ref{fig:Fig1}f. The frequency splitting between the $m_s=|0\rangle \rightarrow |\pm1\rangle$ transitions is given by $\frac{2 g \mu_B B_z}{h}$, where $g\approx2$ is the NV electronic spin $g$-factor, $\mu_B$ is the Bohr magneton, $h$ is Planck's constant, and $B_z$ is the projection of the magnetic field along the NV axis, while the inhomogenously broadened linewidth of the transition is given by $1/T_2^*$. The measured splitting can therefore be used to determine the local dc magnetic field at the NV center (Fig.~\ref{fig:Fig1}e), with the minimum detectable change in magnetic field given by Eq.~\ref{Eq:Sensitivity}. For a single NV center in bulk diamond and with standard readout of the spin state using fluorescence under optical excitation with 532 nm light, $N=1$, $T_x$ is $T_2^*\approx 1~\mu$s, $C\approx 0.05$, and $\gamma=\frac{g \mu_B}{\hbar}\approx2\pi\times2.8\times10^{10}~ \mathrm{T}^{-1}\mathrm{s}^{-1}$. This results in a typical sensitivity to dc magnetic fields of $\eta_{\mathrm{dc}}\approx 0.1~\mu$T$/\sqrt{\mathrm{Hz}}$. An additional figure of merit not captured by Eq.~\ref{Eq:Sensitivity} is the distance from the NV to the diamond surface, which often determines both the spatial resolution of a local measurement and the strength of a signal of interest.

As summarized in Fig.~\ref{fig:Fig2}a, in quantum sensing applications materials challenges manifest themselves through their impacts on the parameters that determine sensitivity, including the optical spin contrast $C$, which is impacted by collection efficiency and NV charge stability, the spin state lifetime and coherence times $T_x$, which are limited by interactions with environmental spins and charges, and the number of NV centers $N$ that can participate in a measurement and distance from the surface or sensing target. Nanoscale quantum sensing therefore requires reliable techniques for robust generation of NV centers close to the diamond surface, which in turn implies a need for well-ordered surface chemistries with low magnetic and electric noise.

\subsection{Quantum networks}
\label{sec:Networks}
Photons enable long distance quantum communication, but they suffer from exponential attenuation. ``Quantum repeater" schemes extend the range of communication \cite{duan2001} by using remote entanglement distribution as a resource for teleportation. Such schemes require a long-lived quantum memory (typically a spin), identical photons across all nodes, and an efficient spin-photon interface. Color centers in diamond can exhibit long spin coherence times and efficient optical transitions, making them excellent candidates for quantum networks. Several key elements of quantum repeaters have been demonstrated using NV \cite{Hensen2015,Kalb2017} and SiV$^{-}$ centers \cite{Bhaskar2020,Sipahigil2016}.

Two key parameters for quantum networks are the entanglement generation rate and the memory time (Fig.~\ref{fig:Fig2}b). Remote entanglement relies on two-photon interference to erase which-path information, and the photons must be coherent and identical in order to interfere. Therefore, the optical linewidth must be Fourier transform-limited, which requires cryogenic temperatures to avoid phonon-induced dephasing \cite{fu2009}. Furthermore, the frequency of emitted photons can depend on the local environment; strain can give rise to static inhomogeneity \cite{olivero2013} and electric field noise leads to fluctuations in the optical transition frequency \cite{Wolters2013}. Therefore, the purity and quality of the host material is crucial.

The entanglement generation rate also depends on the quantum efficiency, total emission rate and collection efficiency. Non-radiative decay reduces the spin-photon entanglement rate. The emission rate and collection efficiency can be improved with photonics, and there have been significant efforts over the past decade to build photonic devices out of heterogeneously integrated layers \cite{Barclay2011} or directly out of diamond \cite{Hausmann2012,Burek2014}.

The memory time is given by the ground state spin coherence. At the cryogenic temperatures necessary for optical coherence, the spin $T_1$ can be over 1 hour for NV centers \cite{Abobeih2018}. Spin coherence is otherwise limited by magnetic noise from $^{13}$C nuclei or paramagnetic impurities (Fig.~\ref{fig:Fig1}b).

For NV centers, achievable entanglement generation rates are limited by their optical properties. A small fraction of emitted photons are at the zero phonon line, leading to an effective quantum efficiency of around 3$\%$ \cite{Barclay2011}. Furthermore, NV centers have a large permanent dipole moment that couples to electric fields, leading to spectral diffusion of the optical transition (Fig.~\ref{fig:growthfabfig}i), which can shift by 1 GHz with ppb fluctuations in the charge environment. This problem is exacerbated by photoionization during optical addressing. Even in high purity materials there is a large distribution of optical linewidths \cite{Chu2014}, and spectral diffusion is particularly problematic in nanofabricated devices  \cite{Ruf2019}. To date, there have been no demonstrations of high atom-photon cooperativity with NV centers because of these limitations.

These challenges have spurred activity in studying alternative color centers, including group IV-vacancy centers \cite{Iwasaki2017,Siyushev2017}, whose nonpolar point symmetry guarantees the lack of a permanent electric dipole. The SiV$^-$ center shows narrow inhomogeneous linewidths, and has been integrated into nanophotonic cavities with record cooperativity \cite{Bhaskar2020}. The SiV$^0$ center also shows similar optical robustness, and additionally exhibits long spin coherence times of 1 second at 10 K \cite{Rose2018a}. There has also been significant activity towards discovery and development of other material systems, including silicon carbide \cite{Bourassa2020}, rare earth ions \cite{Raha2020}, and molecules \cite{bayliss2020}.

\subsection{Quantum algorithms and simulation}
\label{sec:Simulation}

The control of local entanglement among spins is essential for quantum computation and simulation in diamond and most commonly realized between electron spins or between electron and nuclear spins. Electron-electron spin coupling ($\sim 100$~kHz at 10 nm separation) can be used to generate many-body interacting spin systems that involve up to $10^6$ spins \cite{kucsko2018} and has been proposed as a route towards scalable quantum computing \cite{Yao2012}. In the case of electron-nuclear interaction the coupling strength is approximately 1,000 times weaker and therefore nuclear quantum registers are limited to a length-scale of $\sim 2$~nm, with registers involving up to nine nuclear spins demonstrated thus far \cite{bradley2019ten}. 

Local spin-spin entanglement introduces several requirements (Fig.~\ref{fig:Fig2}c), including strong qubit-qubit coupling (achieved by small separations), robustness against decoherence (unwanted paramagnetic impurities and nuclear spins) and disorder (minimizing inhomogeneities due to \textit{e.g.} material strain), and local addressability.

Seminal experiments include entanglement of multiple electronic spins, between different NV centers \cite{dolde2014high} or among dark spins (N$_S^0$) \cite{degen2020entanglement}. However, scaling up to arrays that form a scalable quantum processor is an outstanding materials challenge because of the non-deterministic nature of NV creation and unwanted defects interrupting such an array (Fig.~\ref{fig:growthfabfig}b). A combination of materials-based efforts, e.g. deterministic patterning of NV centers \cite{chen2019laser}, implantation of well defined dark spin chains \cite{bayn2015generation}, and theoretical proposals that address imperfections \cite{Yao2012} are being actively explored. Likewise optical control and readout of individual spins at the nanometer length scale is necessary. Several techniques \cite{grinolds2014subnanometre,rittweger2009sted} have recently demonstrated spin read out and control at the nanoscale, but a robust implementation remains an outstanding challenge. 

Larger dipolar interacting NV ensembles have enabled the study of thermalization dynamics of macroscopic interacting quantum systems \cite{kucsko2018} and driven many-body states \cite{Choi2017}. Control over density, dimensionality, and disorder provide important tuning knobs in these studies and developments in materials science can play an active role in extending this frontier. For example, nuclear spins at the diamond surface have been proposed as a quantum simulation platform \cite{cai2013large}. However, controlling surface termination and unwanted contamination remains an outstanding challenge (Fig.~\ref{fig:surfacesfig}). Importantly, the presence of an optically detectable spin such as the NV center that is magnetically coupled to the interacting spin bath provides the ability to initialize the spin bath as well as provide local readout. 

NV quantum registers have the advantage that individual nuclear spins can possess coherence times exceeding minutes at cryogenic temperature \cite{bradley2019ten} and seconds at room temperature \cite{Maurer2012}, while maintaining fast single and two qubit gate operations. Furthermore, accessing local nuclear memory qubits through an optically detectable NV spin allows for high readout fidelity ($\mathcal{F} > 99\%$) \cite{robledo2011high}. In addition, the implementation of small scale multi-qubit algorithms for quantum information processing \cite{van2012decoherence,waldherr2014quantum}, sensing \cite{lovchinsky2016} and communication \cite{Kalb2017} have been demonstrated. One of the main problems that nuclear spin based quantum registers face is that $^{13}$C spins are stochastically distributed in the diamond lattice, with current material growth techniques offering precise control of the $^{13}$C spin density but not their individual locations. More deterministic control could potentially be achieved through the growth of a $^{13}$C delta-doped layer or with $^{13}$C ion implantation.

\section{Materials challenges in diamond}
\label{sec:MatChallenges}

\subsection{Growth and defect creation}
\label{sec:growth}
\begin{figure}

\includegraphics[width=174mm]{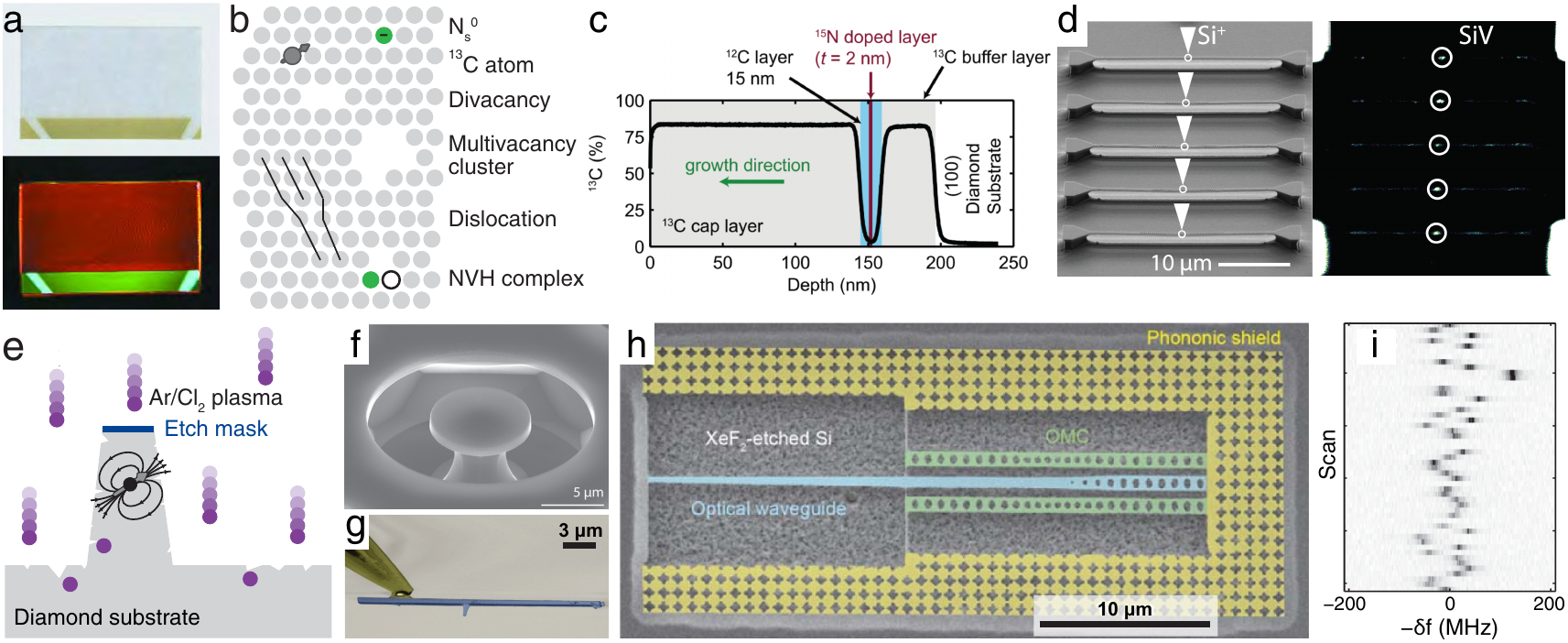}
\caption{\textbf{(a)} Cross-sectional optical image (top panel) of a CVD diamond layer synthesized on a yellow HPHT substrate alongside an ultra-violet excitation-fluorescence image (bottom panel) illustrating red fluorescence in the CVD layer characteristic of NV centers and green fluorescence in the HPHT layer characteristic of extended nitrogen impurities \cite{Martineau2004}. \textbf{(b)}  Cartoon of common defects in bulk diamond resulting from growth and processing. \textbf{(c)} SIMS data demonstrating isotopic layering by CVD growth. This structure was used to extract the 2 nm thickness of a delta-doped layer of NV centers through measurements of the coherent electron-nuclear spin interactions \cite{ohno2012}. \textbf{(d)} SiV$^-$ centers (visible in fluorescence on the right) controllably implanted with a focused ion beam into photonic crystal cavities (shown in a scanning electron microscopy (SEM) image on the left) \cite{Sipahigil2016}. \textbf{(e)} Cartoon showing reactive ion etching of a diamond nanopillar defined via a mask (dark blue at top of pillar). Bombardment from a plasma (here Ar/Cl$_2$ is shown in purple as an illustrative example) leads to implanted ions, and processing can also result in surface roughening. \textbf{(f)} SEM images showing diamond fabricated nanostructures for a variety of applications including a microdisk  \cite{Khanaliloo2015}, \textbf{(g)} scanning probe tip \cite{Appel2016}, and \textbf{(h)} optomechanical crystals \cite{Cady2019}. \textbf{(i)} Resonant excitation of a single NV center in high purity bulk diamond showing spectral diffusion of the zero phonon line \cite{kaimei2009}.}

\label{fig:growthfabfig}
\end{figure}

For all of the applications outlined above, the purity of the host diamond is critical. Bulk diamond growth is challenging because diamond is not the stable allotrope of carbon at atmospheric pressure. Hence, most industrial synthetic single-crystal diamond is synthesized by high pressure high temperature (HPHT) methods. HPHT diamonds typically feature low strain, but they also exhibit high impurity concentrations, typically over 100 ppm nitrogen \cite{Palyanov2010}, which limits NV ensemble coherence times to less than 1 microsecond \cite{Barry2020}. The main synthetic method for quantum applications is plasma-enhanced chemical vapor deposition (CVD), in which diamond is grown homoepitaxially at high temperature in the presence of a carbon-containing gas (typically CH$_4$) and H$_2$ plasmas. High purity CVD substrates are now commercially available with less than 1 ppb nitrogen and boron (Fig.~\ref{fig:growthfabfig}a) \cite{E6}, and isotopic purification up to 99.999\% $^{12}$C has been demonstrated \cite{Teraji2015}.

Beyond the purity of the host material, CVD growth techniques allow for the controlled formation and placement of color centers, enabling and enhancing many of the applications discussed here. An important parameter is the creation efficiency of the color center of interest. The formation of vacancy-related defects in high-quality, as-grown CVD diamond is low and varies widely depending on growth parameters \cite{Barry2020}, so achieving appreciable color center densities typically requires further enhancement with electron irradiation or ion implantation followed by high temperature annealing. These processes introduce other paramagnetic defect species into the lattice such as divacancies or multivacancy clusters (Fig.~\ref{fig:growthfabfig}b), which may contribute to dephasing, charge instability, or inhomogeneous strain environments. Thus, achieving high creation efficiencies while maintaining the high-purity, low-noise environment necessary for long coherence times represents a major materials challenge in this field. 

Precise control over the position of color centers is critical for many quantum applications. CVD offers the possibility of delta doping, which localizes color centers to a depth-confined layer of 1-2 nm thickness \cite{ohno2012}, as illustrated in Fig.~\ref{fig:growthfabfig}c. Another method that offers control over lateral positioning utilizes ion implantation through a nanofabricated mask \cite{Toyli2010} or via a focused ion beam \cite{Sipahigil2016} (Fig.~\ref{fig:growthfabfig}d) but the depth precision is limited to $\sim$10 nm due to ion straggle and channeling effects \cite{Luhmann2018}. Laser writing has also been shown to produce vacancies and allow for three-dimensional positioning \cite{chen2019laser}; however the in-plane accuracy is limited to the average spacing between nitrogen atoms because of vacancy diffusion \cite{chen2019laser}.

Control over a defect's charge state presents yet another challenge. The NV center has two predominant metastable charge states, NV$^-$ and NV$^0$, and ionization events do not preserve spin coherence. Optical control of the NV charge state has been demonstrated \cite{Beha2012,Aslam2013} while other methods of charge state control for NV centers and other defects remains an area of active research. Techniques involving doping diamond with boron or phosphorus have recently been used to pin the Fermi level to stabilize a particular desired charge state of the color center of interest \cite{Rose2018a}. This type of Fermi level engineering may be extended to tune the formation and migration dynamics of other charged species in the diamond lattice as well, allowing for improved control over the surrounding spin environment \cite{FavarodeOliveira2017}.

A broad challenge in optimizing growth is the need for characterizing defects and impurities at the sub-ppb level or in nm-thick layers, while most materials analysis techniques do not have sufficient sensitivity. Secondary ion mass spectrometry (SIMS) allows for detection of heteroatoms at tens of ppb levels \cite{SIMS}, but such information does not include the configuration or charge state of the defect, and the depth resolution can range from a few to hundreds of nanometers depending on analysis conditions \cite{Fiori2014}. Bulk electron spin resonance (ESR) techniques can achieve sensitivities down to $10^{10}$ spins in commercial instruments \cite{bruker} or a few hundreds of spins using microfabricated resonators \cite{Morton2018}; however, this sensitivity scales unfavorably with $T_1$, making it difficult to detect many types of defects. Optical spectroscopy techniques based on absorption or fluorescence can be used to study optically active defects. Generally, many techniques that provide complementary information are required to understand a particular sample. Improving sensitivity of existing techniques and devising new analysis methods would be major enabling tools for improving diamond growth and defect characterization. Furthermore, many of these detection limits may be overcome by using the NV center itself to characterize the surrounding defects, as a complementary spectroscopy tool.

\subsection{Fabrication}
\label{sec:fab}

Building technologies out of color centers in diamond typically requires micro and nanofabrication of the diamond itself. For example, photonic devices can enhance photon collection efficiency and enable coupling to optical or mechanical degrees of freedom. Examples of fabricated structures include nanopillars, scanning probe cantilevers, or nanophotonic and nanomechanical resonators, which are illustrated in Fig.~\ref{fig:growthfabfig}e-h. Many of these structures are fabricated using reactive ion etching or ion milling. Such techniques can induce surface roughness and sub-surface damage, compromising the properties of nearby color centers. The microscopic origins of etch-induced damage are not well-characterized but have been shown to result in spectral diffusion (Fig.~\ref{fig:growthfabfig}i) \cite{Ruf2019}, charge state instability \cite{hauf2011}, reduced fluorescence \cite{Cui2014} and decreased spin coherence \cite{Kim2014}. Ar/Cl$_2$ based etches yield smooth surfaces with roughness below $\sim$ 1 nm rms \cite{Ruf2019,Tao2014}, but etch rates are slow and show poor selectivity, and sub-surface Cl contamination is suspected to contribute to spectral diffusion and spin dephasing, as well as a degradation of quality factors of diamond mechanical resonators \cite{Tao2014}. O$_2$ based etches provide higher selectivity and offer faster etch rates but the combination of micromasking \cite{Ruf2019} and anisotropy can lead to significant surface roughening.

Another major challenge is the lack of high quality heteroepitaxial single-crystal diamond growth, which precludes the isolation of a thin film by conventional undercut processes \cite{Painter1819}. Thin diamond films or device layers with uniform thickness and smooth, minimally damaged surfaces serve as a starting point for a variety of nanophotonic, nanomechanical, and scanning probe devices. Current methods include angled etching \cite{Burek2014}, quasi-isotropic etching \cite{Khanaliloo2015}, focused ion beam (FIB) milling \cite{Maletinsky2012}, brute force thinning via a combination of laser cutting, polishing, and reactive ion etching \cite{Ruf2019,Cady2019,ovartchaiyapong2014} and lift-off techniques using sacrificial ion-damaged layers \cite{lee2014}. Overgrowth of CVD diamond is also an emerging strategy to create high quality, smooth membranes and improve spin properties in fabricated nanostructures  \cite{lee2014}.

\subsection{Surfaces}
\label{sec:surfaces}
\begin{figure}
\includegraphics[width=84mm]{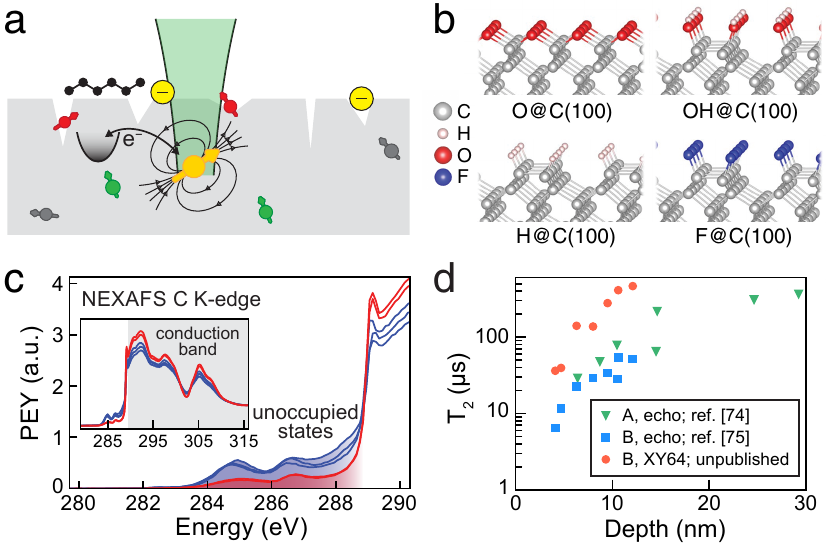}
\caption{\textbf{(a)} Cartoon summary of surface-related materials challenges. The color center (orange) in a host diamond crystal (gray) is shown under green laser illumination. The surface is morphologically rough, contaminated with adventitious hydrocarbons (black), and hosts unwanted charges (yellow) and spins (red), which contribute electric and magnetic field noise respectively. Surface-related charge traps can exchange electrons with the color center, leading to charge instability. The bulk diamond crystal hosts defects from growth and processing such as N$_S^0$ centers (green) and $^{13}$C atoms (grey). \textbf{(b)} Surface bonding configurations for different chemical terminations for (100) oriented diamond surfaces \cite{chou2017}. \textbf{(c)} Carbon k-edge NEXAFS spectra showing trap states associated with the surface for two different surface preparations (red and blue) with multiple samples. The trap state density is suppressed by realizing a more highly ordered oxygen termination (red curve) \cite{Sangtawesin2018}. \textbf{(d)} Hahn echo coherence ($T_2$) times as a function of depth for samples A (green triangles, Ref.~\cite{Myers2014}) and B (blue squares, Ref.~\cite{Sangtawesin2018}). Coherence times for the same NV centers in sample B obtained with XY-64 pulse sequence (red circles) are also shown (previously unpublished).}
\label{fig:surfacesfig}
\end{figure}

Although NV centers in the bulk can have exceptional properties, in general the spin $T_1$ and $T_2$ \cite{ohno2012,FavarodeOliveira2017,Myers2014,Sangtawesin2018}, charge state stability \cite{bluvstein2019,yuan2020}, and ODMR contrast \cite{yuan2020} have been observed to degrade significantly near the diamond surface (Fig.~\ref{fig:surfacesfig}a,d). Surface states in the bandgap arise naturally from the abrupt end to the periodic potential of a crystal. Such states act as electronic traps, giving rise to magnetic and electronic noise because they can be transiently occupied under illumination. The density and electronic position of these states depends sensitively on impurities, structural defects, and the surface chemistry (Fig.~\ref{fig:surfacesfig}b), pointing to the need for careful surface preparation. Diamond surfaces in particular are notoriously difficult to control. Diamond is the hardest known material, making polishing difficult. It is also inert, making wet chemical functionalization challenging. Furthermore, surface processing is often strongly hysteretic and anisotropic, leading to uncontrolled surface morphology \cite{Sangtawesin2018}. Because of these properties, surface processing tends to be highly damaging, leading to strain and extended defects.

Surface morphology has a direct impact on spin coherence, and recent work showed that surface preparation techniques to create smooth, low-defect, oxygen terminated surfaces reduces the density of unoccupied states near the conduction band, leading to NV centers less than 10 nm from the surface with 100 $\mu$s coherence times (Fig.~\ref{fig:surfacesfig}c,d) \cite{Sangtawesin2018}. Separately, implanting nitrogen through a sacrificial boron doped layer can mitigate implantation damage, yielding similar coherence times for shallow NV centers \cite{FavarodeOliveira2017}. An alternative to direct surface engineering is to apply pulse sequences designed to mitigate deleterious surface effects. Dynamical decoupling sequences can further extend shallow NV center coherence \cite{naydenov2011}. Furthermore, recent work has shown NV center coherence can be extended by driving spin defects at the surface  \cite{bluvstein2019extending}. 

Additionally, the charge stability of shallow NV centers degrades with proximity to the surface \cite{bluvstein2019,yuan2020}, reducing ESR contrast \cite{yuan2020} and necessitating experimental checks of the charge state to enhance readout fidelity \cite{bluvstein2019,hopper2020}. The desire to control and stabilize the charge state of shallow NV centers has motivated considerable efforts to control the surface termination of diamond and to understand its impact on defect charge states \cite{hauf2011,cui2013,kaviani2014}.

Despite this recent progress, demonstrated coherence times within less than ten nanometers of the surface are two orders of magnitude lower than the room temperature $T_1$ limit, leaving room for substantial improvement. Future work will likely involve creating surface terminations beyond the currently demonstrated suite of oxygen terminations. Recent work has included investigating fluorine \cite{cui2013} and nitrogen \cite{stacey2015} terminations, although these terminations have yet to improve NV center coherence. One possibility is using hydrogen terminated diamond, which has been shown to yield highly ordered \cite{bobrov2003} and low defect \cite{stacey2019} surfaces but leads to charge state instability of the NV center in ambient conditions because of charge transfer to surface adsorbates \cite{hauf2011}. Cleaning the surface in ultrahigh vacuum conditions could provide a path toward using hydrogen terminated surfaces for NV applications. Furthermore, different surface orientations may allow for access to different chemical terminations; for example chemical groups that are sterically prohibited on the (100) surface may be permitted on (111) surfaces \cite{chou2017}. Such surfaces are difficult to prepare because polishing is more challenging for these orientations, but recent work has shown that near-atomically smooth surfaces are possible with cleaving \cite{parks2018}.

Studying the diamond surface is challenging. Typical surface spectroscopy probes that rely on X-rays (Fig.~\ref{fig:surfacesfig}c), photoelectrons, and electron diffraction are hampered by the highly insulating nature of diamond, carbon background in beamline tools \cite{watts2006}, and ubiquitous carbon-containing surface contaminants \cite{thoms1994}. Furthermore, most of these techniques have sensitivities at the 0.1$\%$ level and integrate over large areas, in contrast to the NV center, which probes its local environment. Addressing this scale mismatch requires a general strategy of correlating surface spectroscopy with NV measurements to evaluate surface processing techniques and discover microscopic sources of noise.

\subsection{Functionalization}
\label{sec:functionalization}

Chemical methods to functionalize the diamond surface could enable direct attachment of molecules to the diamond surface for nanoscale sensing, tunable surface chemistries with desirable properties for shallow color centers, or patterning the surface with interacting or reporter molecules. Diamond is inert and has a small lattice constant, making surface functionalization difficult \cite{szunerits2014}. Typical approaches to control single crystal diamond surface chemistry require harsh conditions such as plasma \cite{hauf2011,cui2013,stacey2015}, thermal annealing \cite{lovchinsky2016,Sangtawesin2018}, or oxidizing acid \cite{Myers2014,stacey2019}. These methods can damage the surface, and limit the scope of functional groups that can be covalently attached to the diamond surface.

Alternatively, wet chemical methods can expand the range of functional groups without damaging the surface. Prior works on functionalizing nanodiamonds have been reported \cite{krueger2017}, which can be used for nanoscale sensing \cite{barton2020} and \textit{in situ} biosensing \cite{Kucsko2013}. Compared to single crystal diamond, nanodiamonds have increased surface area, uncontrolled surface chemistry, and higher levels of defects, all of which introduce undesired electronic states into the band gap \cite{thalassinos2020}. Similarly, micro and polycrystalline surfaces also tend to have disordered chemistry and higher levels of defects \cite{nebel2003}. Such defects provide an interface for functionalization, but also lead to noise and charge traps that degrade NV center properties. Another strategy is to surface coat nanodiamonds with readily functionalizable materials, including inorganic (i.e. SiOx) and organic (i.e. polyglycerol) layers \cite{miller2020}. Most techniques demonstrated with nanodiamonds, micro, or polycrystalline diamonds have not been translated to single crystal samples. One exception is photoelectrochemical and electrochemical methods, which have been demonstrated on single crystal, hydrogen-terminated, diamond substrates \cite{nebel2007}, but these techniques have not been translated to color center applications. Modern catalytic methods \cite{gunawan2014} may provide a new route to control single crystal diamond surface chemistry.

Another challenge is developing a reaction discovery pipeline to accurately and efficiently screen for successful reactions. Many common analysis tools used for solution phase chemistry such as liquid NMR and infrared absorption offer prompt and detailed feedback but are not sensitive enough for surfaces. Instead surface-sensitive analysis tools such as X-ray photoelectron spectroscopy and atomic force microscopy that provide less chemical specificity can be used for rapid feedback, combined with lower-throughput synchrotron techniques. 

\section{Outlook}
\label{sec:outlook}
In order to realize the promise of color centers for applications such as quantum networks, nanoscale NMR, and quantum simulation, there remain a large number of materials challenges to overcome. The sensitivity of defects to their environment sharpens these challenges, but also presents an opportunity to use them as probes of local material properties, opening a new frontier of using quantum probes for nanoscale materials characterization.

Progress towards addressing materials issues in diamond will require a sustained, interdisciplinary, collaborative effort to learn about sources of loss, noise, and decoherence. In particular, close collaborations among physicists, materials scientists and chemists will allow for the systematic study of microscopic sources of noise. These efforts also require advances in theoretical and numerical techniques for predicting the properties of defects and their interactions with the host material \cite{kaviani2014,chou2017,Weber2010,alkauskas2014,Deak2014,Thiering2018,razinkovas2020}. Furthermore, interrogating material properties at the levels required for quantum technologies will require the invention of new processing and characterization tools. Looking forward, some of the materials issues outlined above may be addressed with new defects in diamond that are less sensitive to their environment \cite{Rose2018a,Siyushev2017} and with alternative material platforms, such as SiC \cite{Bourassa2020} and molecular qubits \cite{bayliss2020}. 

We highlight two grand materials challenges that together will make an enormous impact in the development of quantum technologies based on color centers in diamond: first, greater control and understanding of surfaces and interfaces will have broad impact for all color center applications, and will likely also play a large role in any material system used for quantum applications. Second, deterministic placement of color centers at the level of nanometers or even angstroms will allow for control over interactions among qubits and with sensing targets as well as integration into quantum networks. While these represent major technical and scientific hurdles, the rapid progress in the field over the past several decades combined with a recent increase in interest and investment in interdisciplinary quantum science research gives us optimism that they can be overcome, opening the door to the realization and adoption of a wide range of diamond quantum technologies. 

\vskip 0.2in

\noindent \textbf{Acknowledgments}

\noindent We thank Suong Ngyuen for discussions about diamond surface functionalization and Xiaofei Yu for providing the confocal image in Fig.~1g. S.K.~was supported by the U.S. Department of Energy (DOE), Office of Science, Basic Energy Sciences (BES) under Award \# DE-SC0020313.  L.V.H.R.~acknowledges support from the Department of Defense through the National Defense Science and Engineering Graduate Fellowship Program. L.B.H.~acknowledges support from the NSF Quantum
Foundry through Q-AMASE-i program Award No. DMR-1906325. M.X. and P.C.M.~were supported by the National Science Foundation (NSF) Grant No. OMA-1936118. A.C.B.J.~ acknowledges support from the U.S. Department of Energy (DOE) Office of Science, Basic Energy Sciences (BES) under Award \# DE-SC0019241, and by the NSF Grant No. DMR-1810544.  N.P.d.L. was supported by the U.S. Department of Energy (DOE), Office of Science, Basic Energy Sciences (BES) under Award \# DE-SC0018978, and by the NSF under the CAREER program (Grant No. DMR-1752047).

\vskip 0.2in

\noindent
\textbf{Conflict of interest}
 On behalf of all authors, the corresponding author states that there is no conflict of interest.


\bibliographystyle{spphys}       

\bibliography{bibliography.bib}   


\end{document}